\documentclass[a4paper]{spie}  %>>> use this instead for A4 paper
\usepackage[]{graphicx}

\title{Development of a Si/CdTe semiconductor Compton telescope} 

\author{Takaaki Tanaka\supit{a,b}, Takefumi Mitani\supit{a,b}, 
Shin Watanabe\supit{a},  Kazuhiro Nakazawa\supit{a}, 
\\ Kousuke Oonuki\supit{a,b}, Goro Sato\supit{a,b}, 
Tadayuki Takahashi\supit{a,b}, \\Ken'ichi Tamura\supit{a,b}, 
Hiroyasu Tajima\supit{c}, Hidehito Nakamura\supit{d}, 
Masaharu Nomachi\supit{d}, \\Tatsuya Nakamoto\supit{e} 
and Yasushi Fukazawa\supit{e}
\skiplinehalf
\supit{a} Institute of Space and Astronautical Science, JAXA, Sagamihara, Kanagawa, 
229-8510, Japan \\
\supit{b} Department of Physics, University of Tokyo, Bunkyo-ku, Tokyo, 
113-0033, Japan \\
\supit{c} Stanford Linear Accelerator Center, Stanford, CA 94309-4349, 
USA \\
\supit{d} Department of Physics, Osaka University, Toyonaka, Osaka, 
560-0043, Japan \\
\supit{e} Department of Physics, Hiroshima University, Higashi-Hiroshima, 
Hiroshima, 739-8526, Japan
}

\begin{document} 
  \maketitle 

%%%%%%%%%%%%%%%%%%%%%%%%%%%%%%%%%%%%%%%%%%%%%%%%%%%%%%%%%%%%% 
\begin{abstract}
We are developing a Compton telescope based on high resolution 
Si and CdTe imaging devices in order to obtain a high sensitivity astrophysical 
observation in sub-MeV gamma-ray region. In this paper, recent results 
from the prototype Si/CdTe semiconductor Compton telescope are 
reported. The Compton telescope consists of a double-sided Si strip detector (DSSD) 
and CdTe pixel detectors, combined with low noise analog LSI, VA32TA. 
With this detector, we obtained Compton reconstructed images and spectra from 
line gamma-rays ranging from 81~keV up to 356~keV. The energy resolution 
is 3.8~keV and 7.9~keV at 122~keV and 356~keV, respectively, and the angular resolution 
is $9.9^{\circ}$ and $5.7^{\circ}$ at 122~keV and 356~keV, respectively. 
\end{abstract}

%>>>> Include a list of keywords after the abstract 


%%%%%%%%%%%%%%%%%%%%%%%%%%%%%%%%%%%%%%%%%%%%%%%%%%%%%%%%%%%%%
\section{Introduction}
Gamma-ray universe in the energy band of 100~keV to 10 MeV is a 
good observational window to study nucleosynthesis and particle acceleration 
in high energy astrophysical objects. 
However, in this energy band, observation sensitivity have been still limited, 
due to high background, difficulty in detecting photons, and incapability of 
imaging. 
A Compton telescope tracking the Compton scattering process of gamma-rays by 
position and energy sensitive detectors is difficult but a promising approach to 
bring a breakthrough in this energy band. 

In the MeV region, 
COMPTEL\cite{Schoenfelder} on-board {\it CGRO} 
({\it Compton Gamma-Ray Observatory}) was 
the first successful Compton telescope and 
brought many pioneering results\cite{Schoenfelder2}. 
In a Compton telescope, deposited energy and 
position of the gamma-ray interactions with the detector are acquired.  
When a gamma-ray photon is scattered in one detector and absorbed in another detector,  
the incident energy of the gamma-ray and the scattering angle can be determined as, 
\begin{eqnarray}
E_{\rm in} &=& E_1 + E_2\label{eq:comp_ene}\\
\cos\theta &=& 1 - \frac{m_e c^2}{E_2} + \frac{m_e c^2}{E_1 + E_2}\label{eq:comp},
\end{eqnarray}
where $E_1$ is the energy of the recoil electron, $E_2$ is the energy of 
the scattered photon and $\theta$ is the scattering angle. 
The latter equation presents the incident direction as a cone in the sky. 

As shown in equation (\ref{eq:comp}), incident direction of a gamma-ray is calculated 
from the position and the energy information of interactions. Therefore, employing 
semiconductor imaging spectrometers as components of Compton telescopes, 
with their good energy and position resolution, will drastically improve the angular 
resolution and hence the sensitivity. In this point of view, advanced Compton telescopes 
based on Si, Ge, CZT and CdTe are proposed and development is on-going in many 
groups\cite{Takahashi04NAR,Takahashi03,MEGA,Kurfess,TIGRE,CZT}. 
A Compton telescope also enables us to measure polarization of 
gamma-rays, by obtaining the azimuthal distribution of 
Compton scattering\cite{Kamae,polarimetry}. 

Based on our CdTe and Si detector technologies\cite{Takahashi02,Tanaka,Nakazawa,Fukazawa,Tajima03}, 
we are working on a new Si/CdTe semiconductor Compton telescope 
to explore the universe in the energy band from 100~keV to a few MeV. 
In this paper, we describe  experimental results 
obtained with a prototype Compton telescope which consists of 
one layer of a Double-sided Si strip detector (DSSD) and one layer of CdTe pixel 
detectors.  Detailed performance of the DSSD and the vision of next-generation 
gamma-ray missions using this technique are reported elsewhere\cite{Fukazawa,Takahashi04}.
Demonstration of polarization detection are reported 
in Mitani {\em et al.}\cite{Mitani04} and Tajima {\em et al.}\cite{Tajima04}

\section{Si/CdTe semiconductor Compton Telescope}
Basic concept of Si/CdTe Compton telescope is
to utilize Si as a scatterer and CdTe as an 
absorber\cite{Takahashi03,Mitani04}.
Compared to Compton telescopes using scintillator as an absorber,
this design improves the performance 
especially at sub-MeV region, as low as 100 keV.
The lower energy coverage is prefered to connect the band pass 
of Compton telescope with that of the technology using 
hard X-ray focusing mirror optics\cite{Yamashita,HEFT}.

\begin{figure}[htbp]
\begin{center}
\includegraphics[width=.35\linewidth,clip]{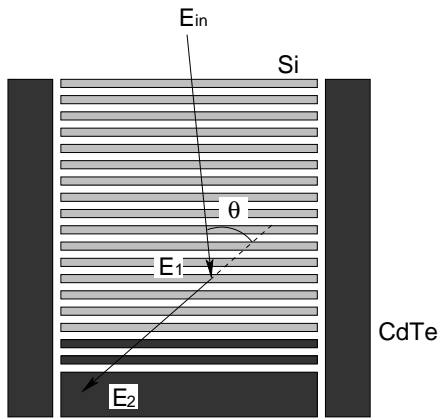}
\caption{Schematic picture of our Si/CdTe semiconductor Compton telescope 
under development}
\label{fig:si_cdte_comp}
\end{center}
\end{figure}

\begin{figure}[htbp]
\begin{center}
\includegraphics[width=.5\linewidth,clip]{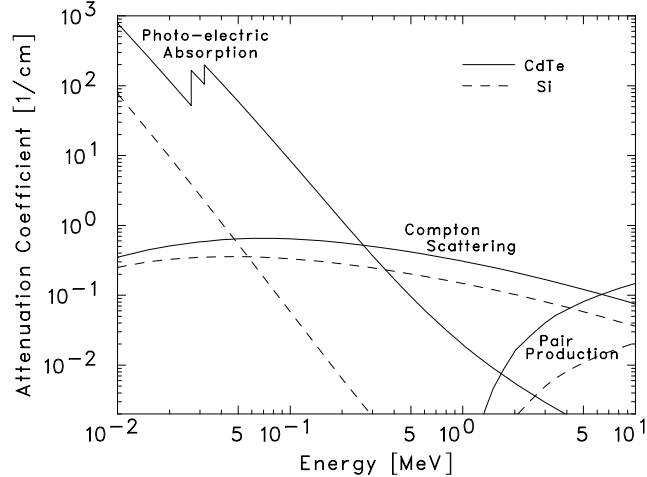}
\caption{Linear attenuation coefficients for photo-electric absorption, Compton scattering 
and pair production in Si and CdTe.}
\label{fig:tau}
\end{center}
\end{figure}

A schematic image of Si/CdTe Compton telescope is shown 
in Figure \ref{fig:si_cdte_comp}.
The major interaction process of Si 
with gamma-rays with energy higher than 60 keV is Compton scattering,
while in CdTe photo-electric absorbtion is still dominant up to 300 keV
(Figure \ref{fig:tau}).
In a Si/CdTe Compton telescope, therefore, ``one-Compton and absorption'' 
events dominate the data around several hundred keV.
A merit of utilizing this kind of event is that it is relatively easy to analyze.
For example, in Compton scattering of photons with energy less than
$m_e/2 = 255.5$~keV, the recoil electron energy is 
always less than that of the scattered photon\cite{Takahashi04NAR}.
In addition, the ambiguity in the Compton scattering angle
caused by the orbital motion of bound electron 
is relatively small for Si compared to that of Ge and CdTe.
This effect is called ``Doppler broadening effect\cite{Zoglauer}'', 
and is the dominant factor of angular resolution around
a few hundered keV.

The demerit of current Si/CdTe Compton telescope design is 
that it looses effective area especially at the higher energy band,
as much as a few MeV.
This is simply because the density of Si ($2.3~{\rm g/cm^3}$) 
is much less than that of CdTe ($5.9~{\rm g/cm^3}$).
For Compton telescope aiming mainly at a few MeV band,
increasing the number and thickness of CdTe layers becomes important.

\section{Experimental setup}
Two key technologies for our Si/CdTe Compton telescope are DSSD and 
CdTe pixel detectors. 
In this section, we describe the detail of the newly developed $8 \times 8$ 
CdTe pixel detector and design of the prototype Compton telescope. 
Detail of the DSSD development is described in Fukazawa {\em et al.}\cite{Fukazawa}

\subsection{The $8 \times 8$ CdTe Pixel Detector}
\begin{figure}[htbp]
\begin{center}
\includegraphics[width=.3\linewidth,clip]{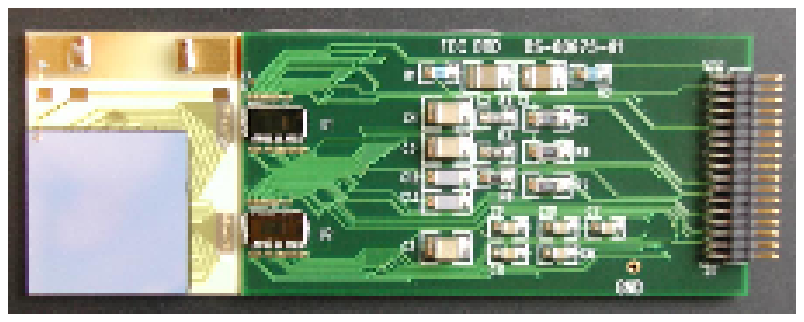}
\includegraphics[width=.45\linewidth,clip]{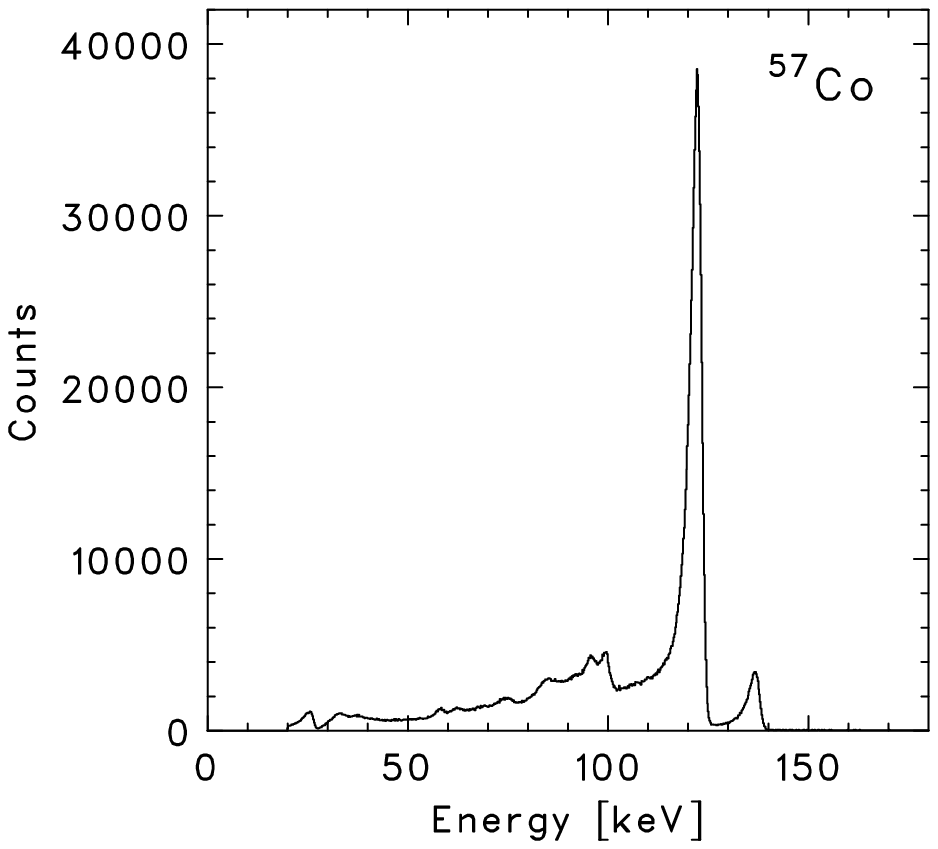}
\caption{Photo of the $8 \times 8$ CdTe pixel detector (left) and ${}^{57}{\rm Co}$ spectrum 
from all 128 channels with the two detectors employed in the prototype Compton 
telescope(right). The operated temperature is $-10~{\rm {}^{\circ}C}$ and the applied bias 
voltage is 500~V. The peaking time of the shaping amplifier in VA32TA is set at 
$2~{\rm \mu sec}$.}
\label{fig:8x8_photo}
\end{center}
\end{figure}

The $8 \times 8$ CdTe pixel detector is based on the Schottky CdTe diode 
device, utilizing indium as the anode and platinum as the cathode. 
This device is developed with ACRORAD, Japan and 
characterized by its low leakage current and 
high uniformity\cite{Takahashi02,Nakazawa03}.
Figure~\ref{fig:8x8_photo} (left) shows the photo of the detector. 
The detector has dimensions of $18.55~{\rm mm} \times 18.55~{\rm mm}$ 
and a thickness of $500~\rm{\mu m}$. The indium side is 
used as a common electrode and the platinum side is 
divided into 8 by 8 pixels and a guard ring with a width of 1~mm. 
The guard-ring electrode are attached to reduce the leakage current 
of the side and corner pixels because most of the leakage current is 
through the detector perimeter\cite{Tanaka,Nakazawa}. 
Pixel size is $2~{\rm mm} \times 2~{\rm mm}$, and the gap between each 
pixel is $50~\rm{\mu m}$. 

Each pixel is connected to a fanout board by bump bonding technology 
developed in cooperation with Mitsubishi Heavy Industry, 
Japan\cite{Takahashi01}.
Since the leakage current of each pixel is as low as 
$\sim 100~{\rm pA}$ at a bias voltage of 500~V even at room temperature, 
the signal lines are directly connected to the input of 
readout electronics, an analog LSI VA32TA\cite{Tajima03,Tajima_ieee04}. 
This LSI is developed with IDEA, Norway, and characterized by its low noise. 
The LSIs are read out via specially designed compact readout system including 
an ADC and FPGA, which is controlled with a fast serial interface 
"Space Wire (IEEE 1355)"\cite{Mitani03}. 

Figure \ref{fig:8x8_photo} (right) shows a spectrum of ${}^{57}{\rm Co}$ 
obtained with the CdTe pixel detectors employed in the prototype Compton 
telescope. In these detectors, all 128 channels are properly connected and their 
spectra are summed after gain correction. 
The data is measured at a temperature of $-10~{\rm {}^{\circ}C}$ and with 
a bias voltage of 500~V. The energy resolution is 3.2~keV (FWHM) for the 
122~keV peak. 

\subsection{The prototype Compton Telescope System}
\begin{figure}[htbp]
\begin{center}
\includegraphics[width=.3\linewidth,clip]{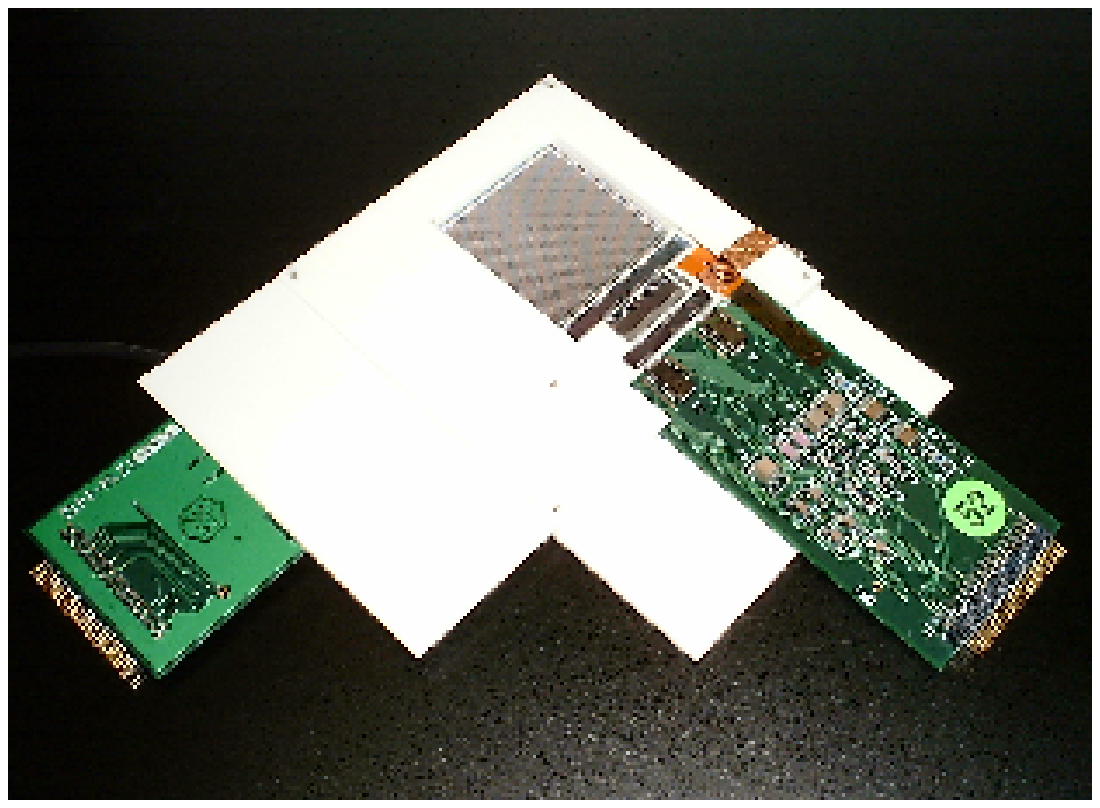}
\includegraphics[width=.45\linewidth,clip]{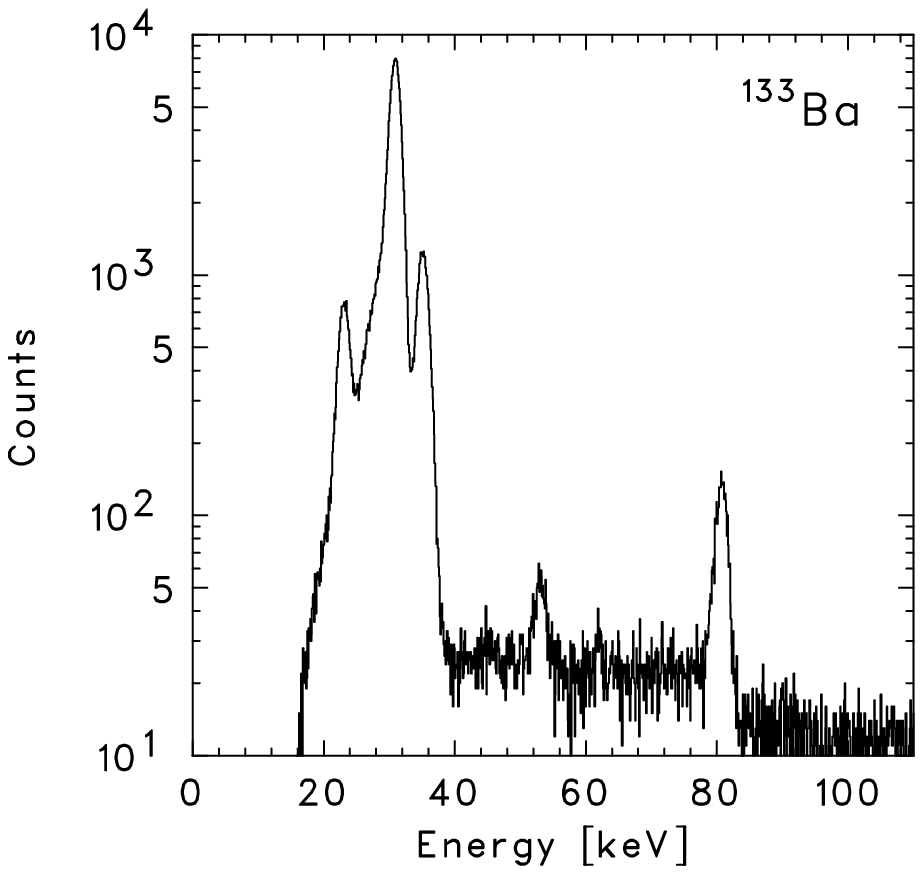}
\caption{Photo of the DSSD (left) and ${}^{133}{\rm Ba}$ spectrum obtained with 
the p-strips at a temperature of $-10~{\rm {}^{\circ}C}$ (right). The peaking time 
of the shaping amplifier is $2~{\rm \mu sec}$.}
\label{fig:dssd_photo}
\end{center}
\end{figure}

The prototype Si/CdTe Compton telescope consists of two CdTe pixel 
detectors and a DSSD. The DSSD has a thickness of 
$300~\rm{\mu m}$. There are 64 strips with a length of 2.53~cm in 
both sides oriented orthogonally. 
The strip pitch is $400~\rm{\mu m}$ and the gap between 
the strips is $100~\rm{\mu m}$. By reading out both the p-strips and 
n-strips, we can obtain two-dimensional positional information. 
The device is also read out with VA32TA. 
The p-strips are connected directly to the input of the LSI, and the n-strips are 
connected via coupling capacitors. To fully deplete the device, 
the n-side is biased with a voltage of 110~V. 
Because the spectral resolution of the p-side is better than that of the n-side, 
we obtained spectra only from the p-strips, and data from the n-strips 
are used for the positional information. 
Figure~\ref{fig:dssd_photo} shows a spectrum of ${}^{133}{\rm Ba}$ 
obtained with the p-strips of the DSSD operated 
at a temperature of $-10~{\rm {}^{\circ}C}$. The energy 
resolution for 81.0~keV peak is 1.9~keV (FWHM). 

\begin{figure}[htbp]
\begin{center}
\includegraphics[width=.4\linewidth,clip]{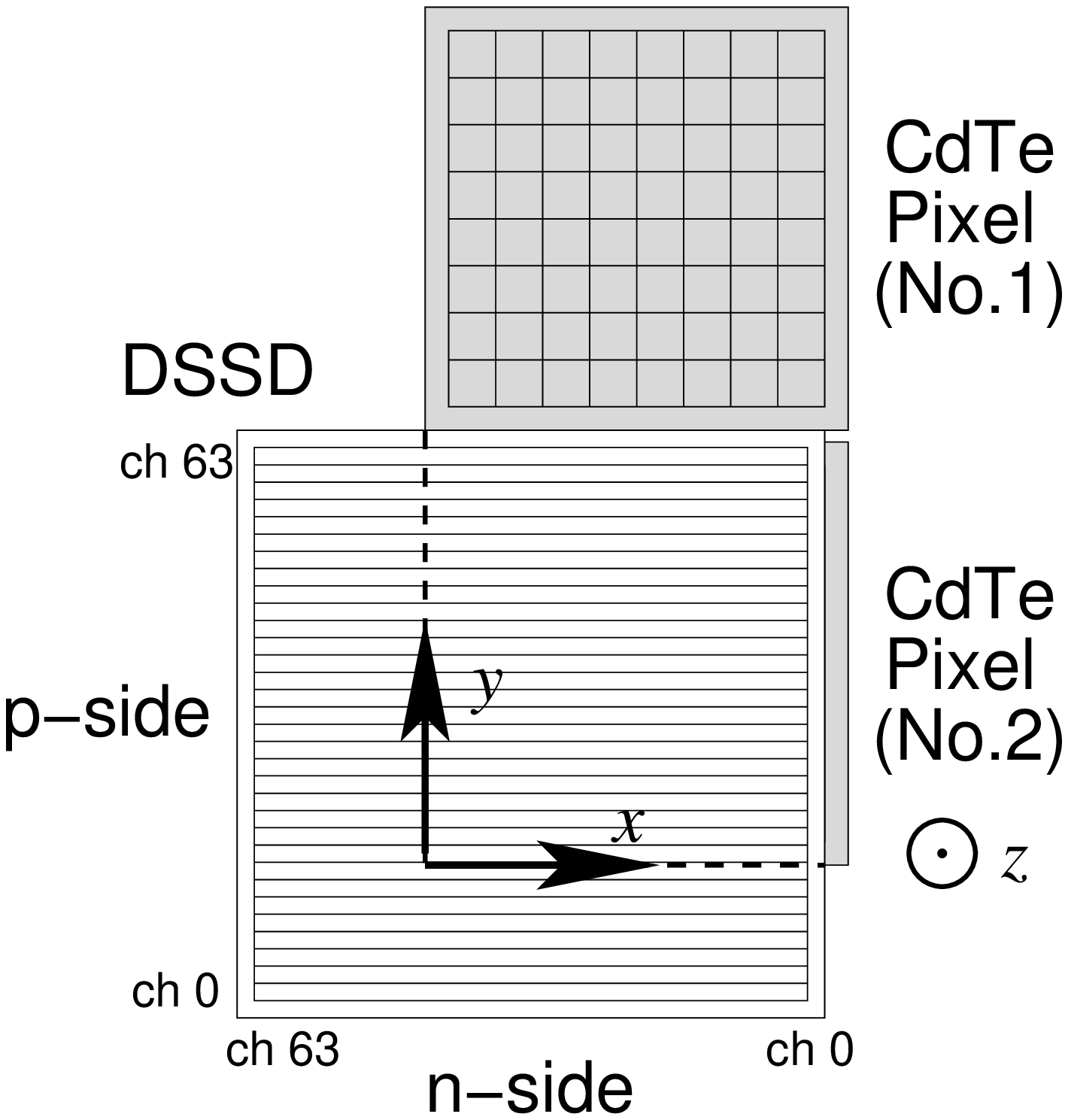}
\includegraphics[width=.4\linewidth,clip]{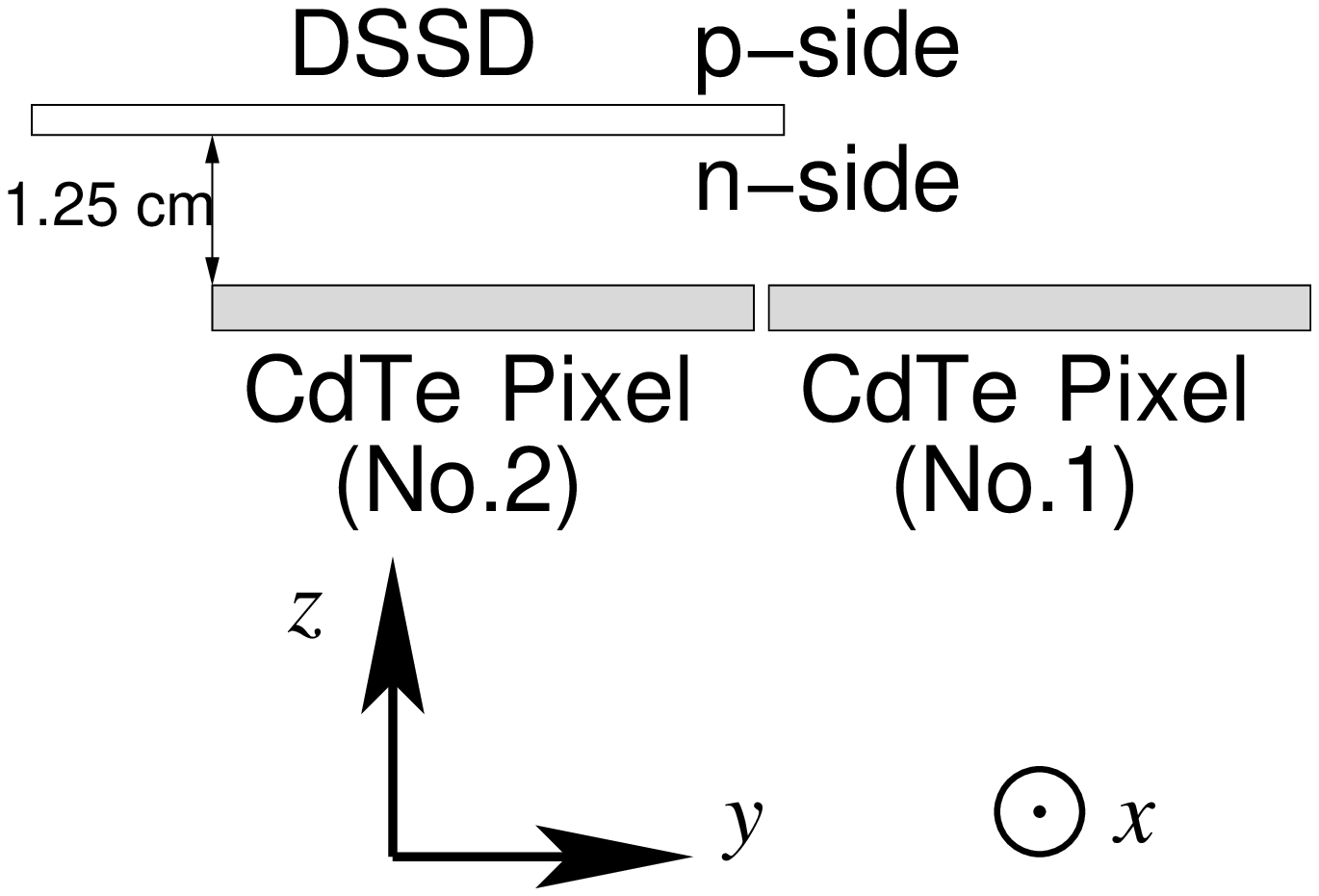}
\caption{Configuration of the prototype Si/CdTe Compton telescope.}
\label{fig:compton_det_config}
\end{center}
\end{figure}

Figure~\ref{fig:compton_det_config} presents the arrangement of two CdTe 
pixel detectors and the DSSD at our Compton telescope. The separation 
between the DSSD layer and the CdTe layer is 12.5~mm. 
In the following experiment, 
we kept the whole detector system at a temperature of 
$-10~{\rm {}^{\circ}C}$. 
We irradiated gamma-ray sources of ${}^{57}{\rm Co}$ and ${}^{133}{\rm Ba}$  
which were placed 353~mm above the surface of the DSSD. 
Triggers only from the LSIs connected to the CdTe detectors were enabled.  
When a trigger was generated from some channel, all channels connected to 
the CdTe detector which issued the trigger and all channels connected to the DSSD 
are read out. 

\section{Data analysis}
\subsection{Compton Reconstruction of Images}\label{subsec:image}
Data reduction is performed as follows. First,``two-hit events'', one hit in DSSD 
and one hit in CdTe, were selected from the raw data. 
Here, one hit in DSSD means that only one channel that is connected to 
DSSD has a pulse hight above 6~keV, and one hit in CdTe 
detectors means that only one channel that is connected to CdTe detectors 
has a pulse hight above 20~keV. 

\begin{figure}[htbp]
\begin{center}
\includegraphics[width=.45\linewidth,clip]{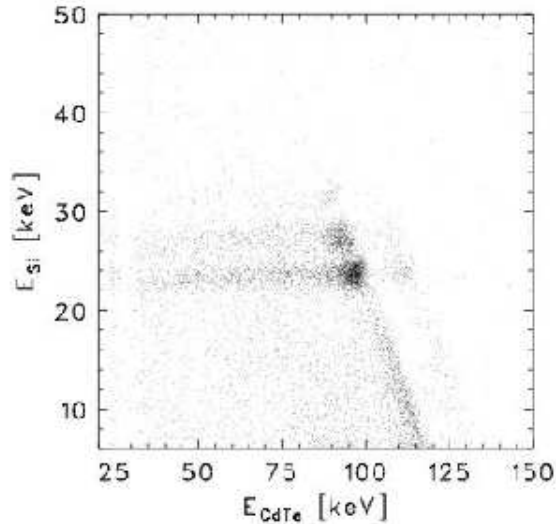}
\caption{Scatter plot of deposited energy at the DSSD ($E_{\rm Si}$) and 
at the CdTe detectors ($E_{\rm CdTe}$) for all two-hit events when we irradiated 
${}^{57}{\rm Co}$ to the prototype detector.}
\label{fig:2dim_plot}
\end{center}
\end{figure}

A scatter plot of the energy deposit at the DSSD ($E_{\rm Si}$) and at the CdTe 
detectors ($E_{\rm CdTe}$) is shown in Figure~\ref{fig:2dim_plot}. This plot 
is obtained when we irradiated ${}^{57}{\rm Co}$ to the Compton telescope. 
There are branches which satisfies $E_{\rm Si}+E_{\rm CdTe}\cong122~{\rm keV}$ and 
$E_{\rm Si}+E_{\rm CdTe}\cong136~{\rm keV}$. These branches are the 
events in which incident gamma-rays of 122~keV or136~keV are scattered 
in one detector and then absorbed in the other. 
Another noticeable branches are those parallel to the horizontal axis 
around $E_{\rm Si} = 20$--$30~{\rm keV}$. 
These are the events by the fuorescence X-rays from 
Cd (${\rm K_\alpha}$: 23.1~keV, ${\rm K_\beta}$: 26.1~keV)
and Te (${\rm K_\alpha}$: 27.4~keV, ${\rm K_\beta}$: 31.0~keV)
which escape from 
the CdTe detectors and are absorbed in the DSSD, and not the Compton events.

\begin{figure}[htbp]
\begin{center}
\includegraphics[width=.48\linewidth,clip]{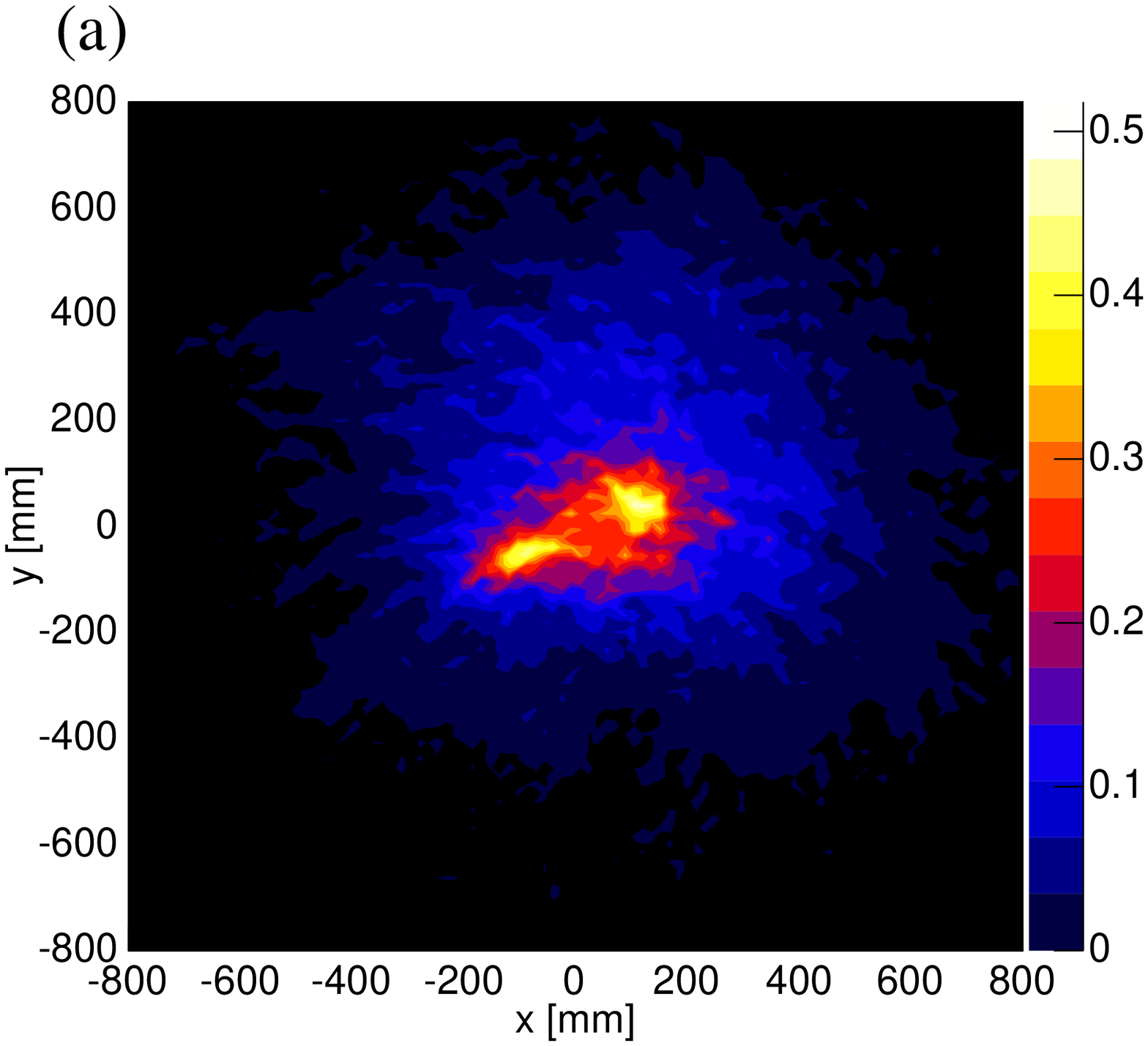}
\includegraphics[width=.48\linewidth,clip]{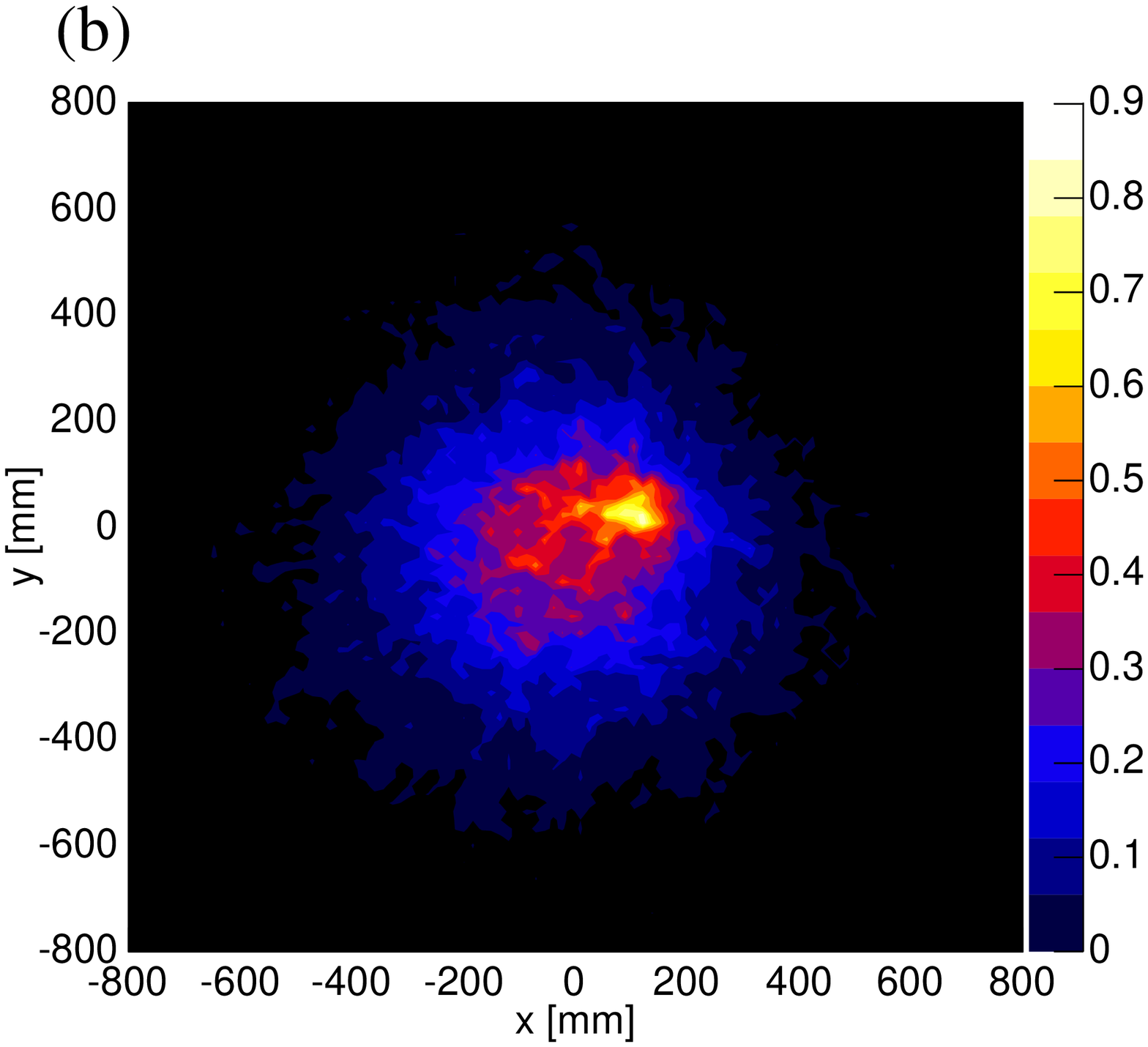}
\includegraphics[width=.48\linewidth,clip]{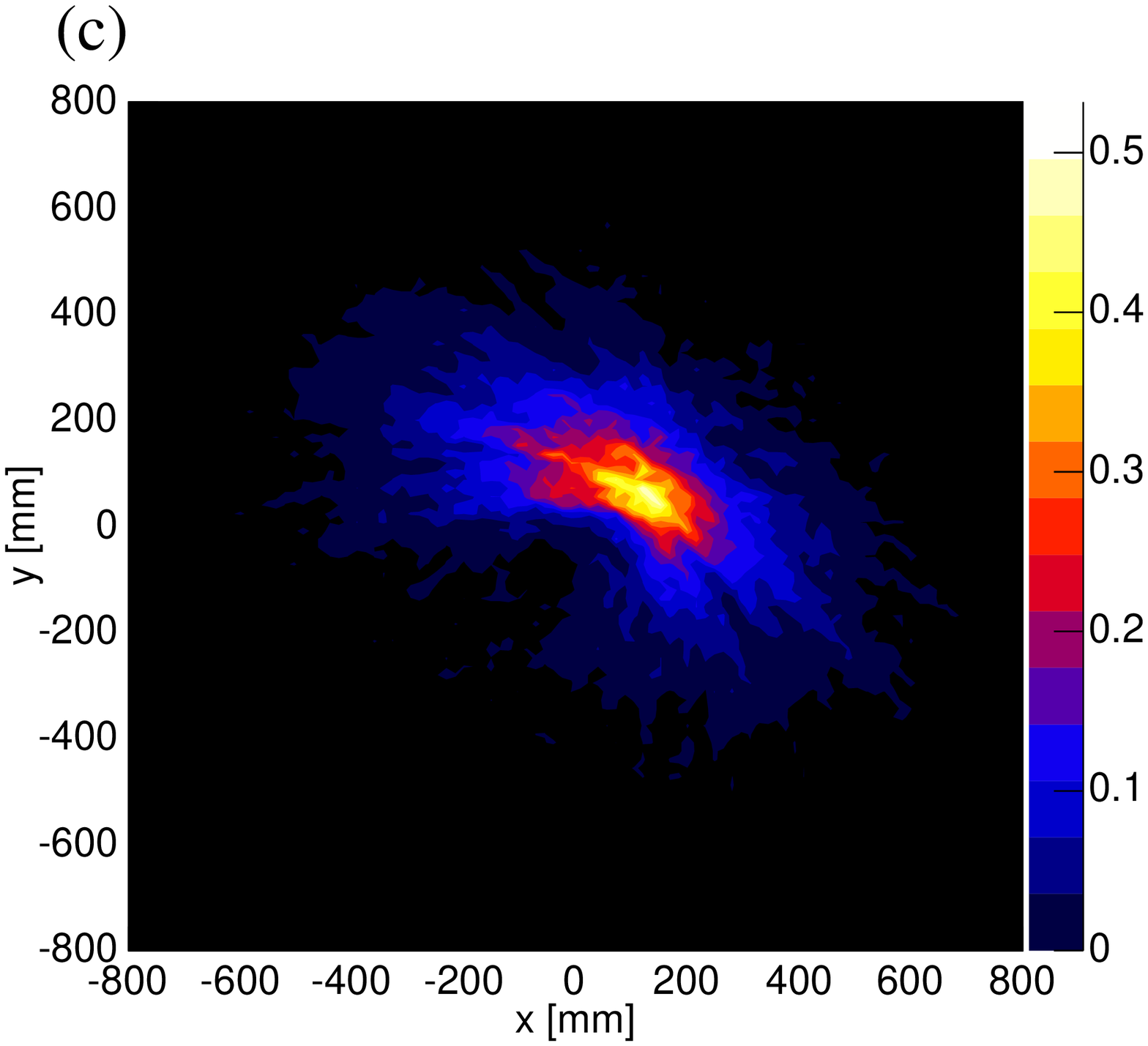}
\includegraphics[width=.48\linewidth,clip]{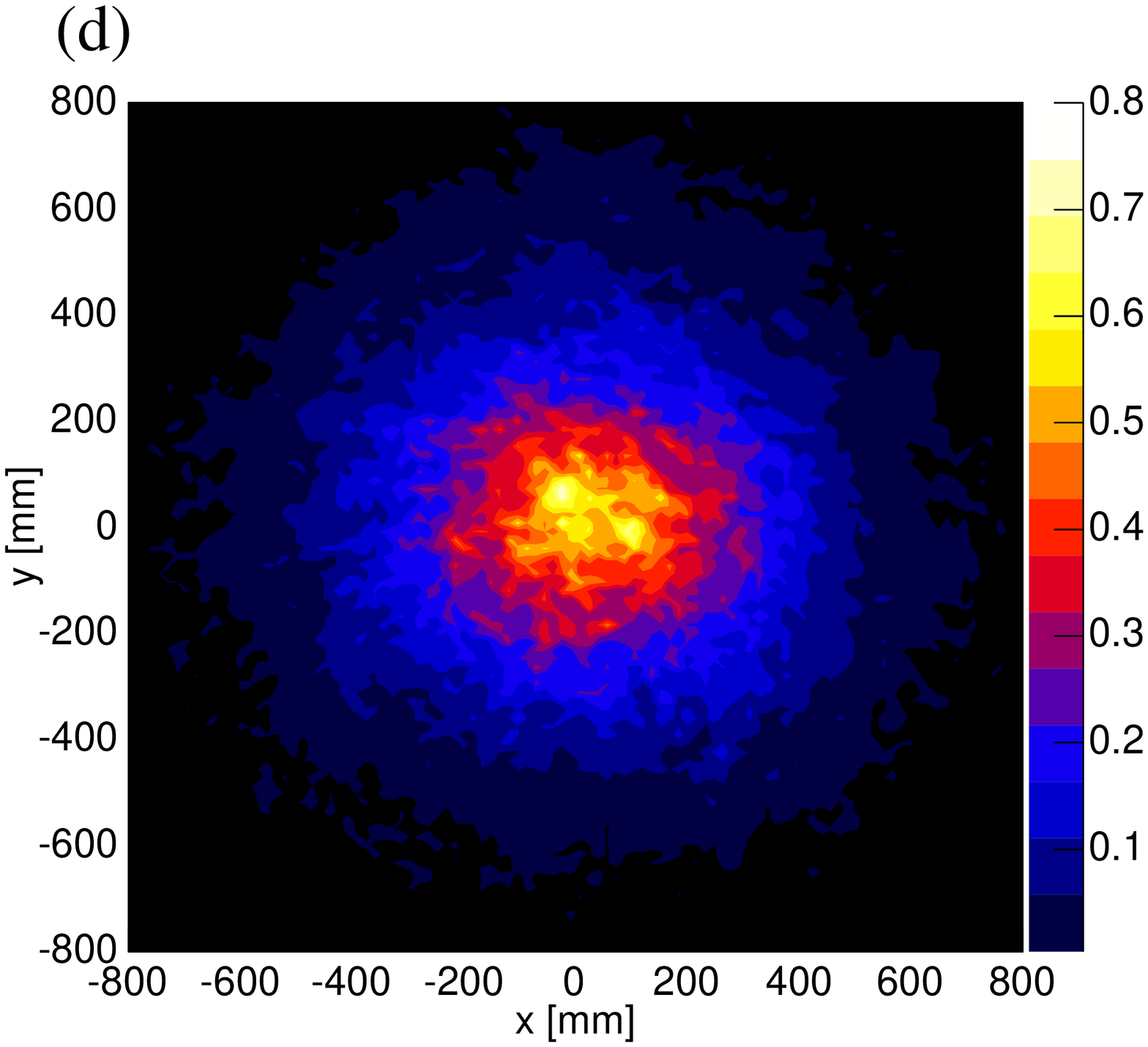}
\caption{Compton reconstructed images. (a) Image of two ${}^{57}{\rm Co}$ sources by events within $122\pm5~{\rm keV}$. (b)--(d) Image obtained by irradiating ${}^{133}{\rm Ba}$. 
(b) Events in which total energy deposit is above 250~keV. (c) Events within $81\pm5~{\rm keV}$. (d) Events from 100~keV to 250~keV.}
\label{fig:comp_image}
\end{center}
\end{figure}

For these ``two-hit events'', we calculated the 
Compton-scattering angle by using equation (\ref{eq:comp}). 
In the calculation, we assumed that 
incident gamma-rays are scattered in the DSSD and absorbed in the CdTe pixel 
detector. From the calculated scattering angle and the two hit positions, the Compton 
cones are drawn on the sky event by event. We projected the cone at the plane 
at the distance of 353~mm, and obtained the image of the gamma-ray source. 
Figure \ref{fig:comp_image} shows the images obtained with 
the experimental data, in which  the gamma-ray source position can be clearly identified. 

Figure \ref{fig:comp_image}(a) is obtained by irradiating gamma-rays from two 
${}^{57}{\rm Co}$ sources. 
These sources are placed 230~mm separate from each other, which 
corresponds to $33^{\circ}$ at the distance of 363~mm.
The ${}^{57}{\rm Co}$ image is obtained by selecting events in which sum of the energy 
deposit at the DSSD and the CdTe detectro is within $122\pm5~{\rm keV}$. 

For the ${}^{133}{\rm Ba}$ images, energy selection is above 250~keV and 
$81\pm5~{\rm keV}$ for Figure \ref{fig:comp_image}(b) and (c), respectively. 
In Figure \ref{fig:comp_image}, appearance of each image is different. This reflects 
the basic process and geometry of the detector. For example, 81~keV image has 
a peculiar ``batterfly-like'' shape. 
This is caused by the fact that the scattering angle must be larger than $60^{\circ}$ 
to deposit an energy larger than 6~keV at the DSSD, which is the energy threshold in 
this analysis, and that Compton cones with peculiar orientation are obtained 
because the CdTe detectors are offset from the DSSD. 
Another impressive image is obtained when we select the energy band of 
100--250~keV for ${}^{133}{\rm Ba}$ irradiation data (Figure \ref{fig:comp_image}(d)). 
Compared with line images (Figure \ref{fig:comp_image}(b) and (c)), the location of 
${}^{133}{\rm Ba}$ source is not clear. This is because the image is dominated by 
back scattered events and ``Compton-Compton'' events which finally escaped from the 
detector. From this data, we see that it is better to use thicker CdTe detector 
and, if possible, utilize a surrounding shield counter to identify the physical process 
of each event.

\subsection{Angular Resolution}\label{subsec:angle_res}
\begin{figure}[htbp]
\begin{minipage}{.5\linewidth}
\begin{center}
\includegraphics[width=.9\linewidth,clip]{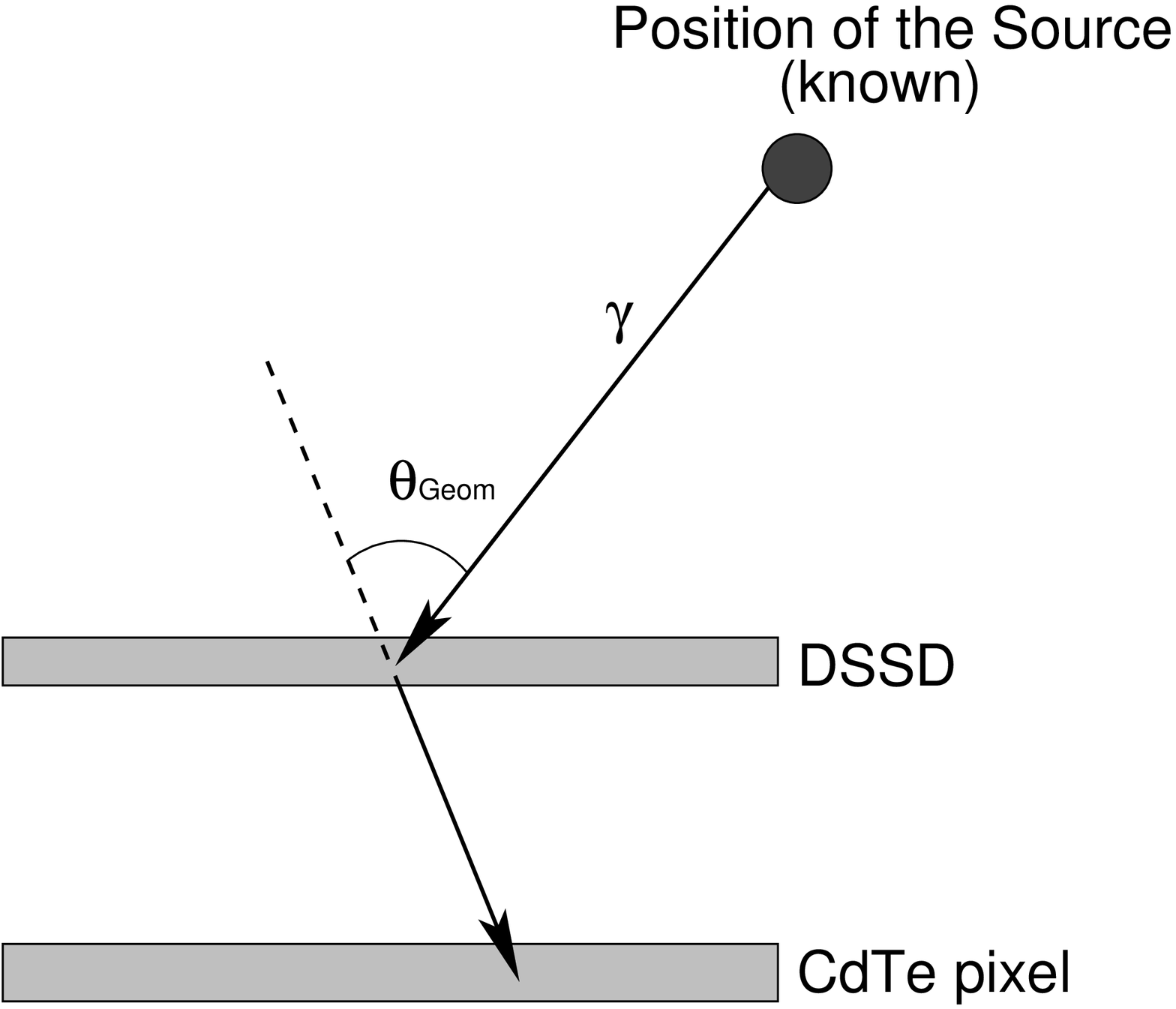}
\caption{Definition of $\theta_{\rm Geom}$}
\label{fig:theta_geom}
\end{center}
\end{minipage}
\begin{minipage}{.5\linewidth}
\begin{center}
\includegraphics[width=.9\linewidth,clip]{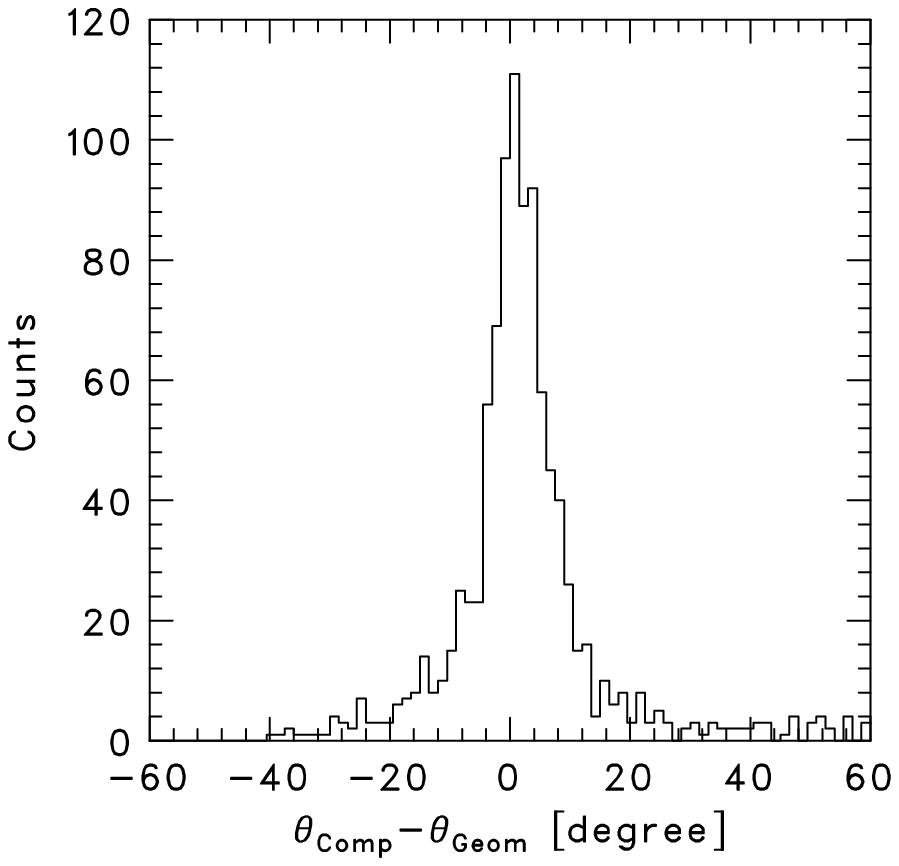}
\caption{Distribution of $\theta_{\rm Comp}-\theta_{\rm Geom}$ for 122~keV events.}
\label{fig:theta_dist}
\end{center}
\end{minipage}
\end{figure}

In order to evaluate angular resolution of our prototype Compton telescope, 
we compare the calculated Compton-scattering angles ($\theta_{\rm Comp}$)
with those defined by the location of gamma-ray source and 
hit positions ($\theta_{\rm Geom}$) (see Figure \ref{fig:theta_geom}). 
The difference between the two values reflects the angular resolution 
of Compton telescope. 

\begin{figure}[htbp]
\begin{center}
\includegraphics[width=.5\linewidth,clip]{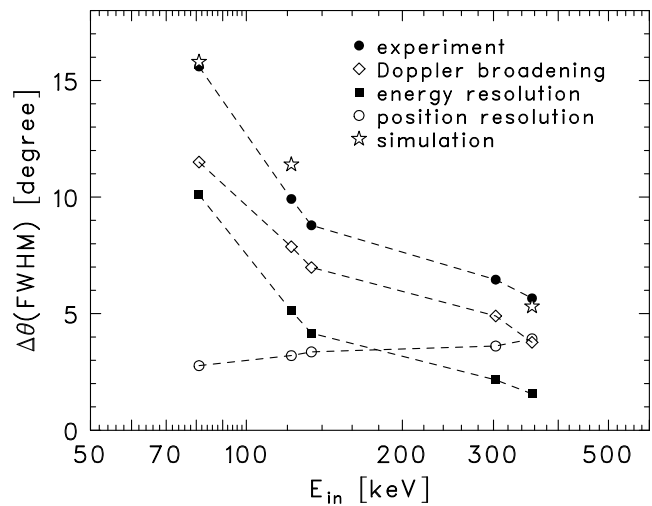}
\caption{Relation between incident gamma-ray energy and angular resolution.}
\label{fig:Ein_dTheta}
\end{center}
\end{figure}

We examined the distribution of $\theta_{\rm Comp} - \theta_{\rm Geom}$ 
for various peaks of ${}^{57}{\rm Co}$ (122~keV and 136~keV)
and ${}^{133}{\rm Ba}$ (81.0~keV, 303~keV and 356~keV). Figure 
\ref{fig:theta_dist} shows the example of 122~keV events. 
The FWHM of these distributions ($\Delta \theta_{\rm exp}$), 
which is the angular resolution of the Compton telescope, is 
plotted against the incident gamma-ray energy in Figure \ref{fig:Ein_dTheta}. 
The angular resolution obtained from Geant4 Monte Carlo simulations\cite{Geant4}, 
using the Compton scattering process, G4LECS\cite{Kippen,GLECS} 
are also plotted. The experimental results are 
consistent with the results from the simulation. 

As shown in Figure \ref{fig:Ein_dTheta}, the angular resolution becomes better 
as the incident gamma-ray energy becomes higher. 
For example, $\Delta \theta_{\rm exp} = 16^{\circ}$ at 81.0~keV and 
$\Delta \theta_{\rm exp} = 5.7^{\circ}$ at 356~keV. 
In order to examine the reason of this tendency, we estimated the contribution
of position and energy resolution of the DSSD and the CdTe detector. 
Contribution of position resolution ($\Delta \theta_{\rm pos}$) 
is estimated by using the actual hit positions obtained from the experiment 
and the strip/pixel size of the DSSD and the CdTe detector. 
To estimate the contribution of energy resolution ($\Delta \theta_{\rm ene}$), 
we used the Compton-scattering angles of actual events and smoothed them 
with detector energy resolution. 
In Figure \ref {fig:Ein_dTheta}, the estimated values are plotted. 
In the same Figure, the values of 
\begin{eqnarray}
\Delta \theta_{\rm DB} = \sqrt{(\Delta \theta_{\rm exp})^2 - 
(\Delta \theta_{\rm pos})^2 - (\Delta \theta_{\rm ene})^2}
\end{eqnarray}
are also plotted. This can be interpreted to be the contribution of Doppler broadening, 
which is supported by the fact that the experimental data is consistent with 
those obtained from simulations. 
As seen in Figure \ref{fig:Ein_dTheta}, $\Delta \theta_{\rm ene}$ and 
$\Delta \theta_{\rm DB}$ becomes smaller as the incident energy becomes higher, 
whereas $\Delta \theta_{\rm pos}$ becomes larger. The tendency of 
$\Delta \theta_{\rm DB}$ is consistent with that of Monte Carlo simulations in 
several papers (e.g. Zoglauer and Kanbach 2003\cite{Zoglauer}). 
$\Delta \theta_{\rm pos}$ becomes 
smaller when the incident gamma-ray energy is smaller, because in such events 
only larger scattering angle and hence the longer distance events 
can be reconstructed due to the energy threshold 
for the DSSD (6~keV). 

\begin{figure}[thbp]
\begin{center}
\includegraphics[width=.5\linewidth,clip]{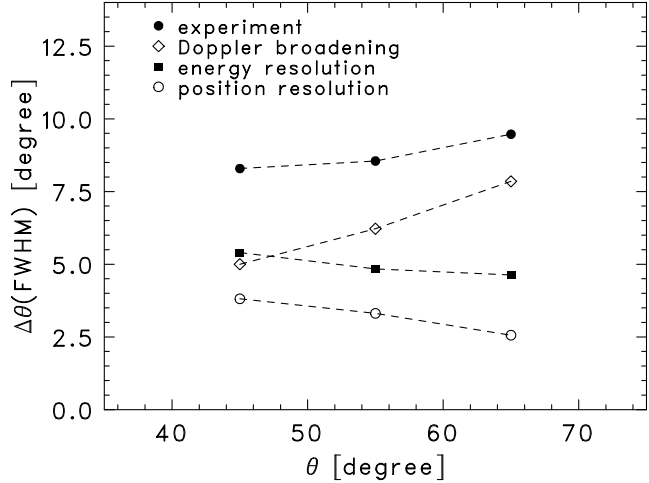}
\caption{Relation between Compton-scattering angle and angular resolution for 122~keV events.}
\label{fig:Theta_dTheta}
\end{center}
\end{figure}

 In order to investigate the relation between the scattering angles and the angular 
resolution, we plotted the angular resolution against Compton scattering angle 
from ${}^{57}{\rm Co}$ data in Figure \ref{fig:Theta_dTheta}. 
The estimated value of uncertainty due to 
position resolution, energy resolution and Doppler broadening are obtained 
in the same way as Figure \ref{fig:Ein_dTheta} and plotted together. 
The value of $\Delta \theta_{\rm exp}$ is somewhat larger for larger scattering 
angles, and our analysis shows that this tendency is caused by 
Doppler broadening. In other words, the contribution of Doppler broadening 
can be reduced if we can reconstruct the events with smaller scattering angles, 
which means lower energy threshold for the DSSD. 

\subsection{Compton Reconstruction of Spectra}
\begin{figure}[htbp]
\begin{center}
\includegraphics[width=.45\linewidth,clip]{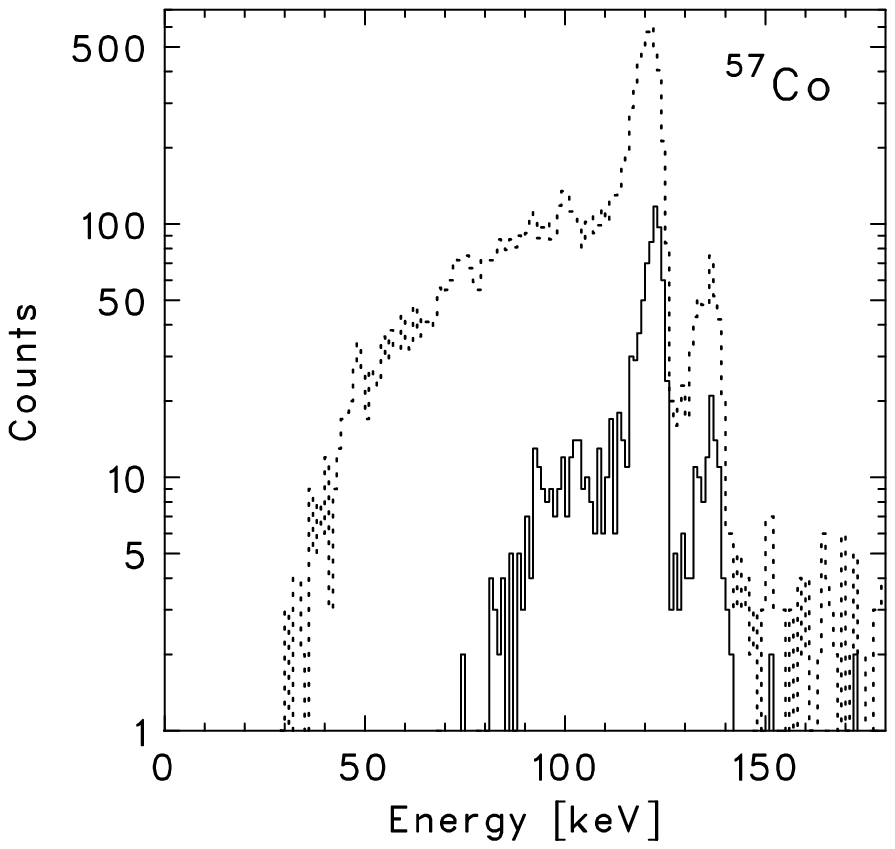}
\includegraphics[width=.45\linewidth,clip]{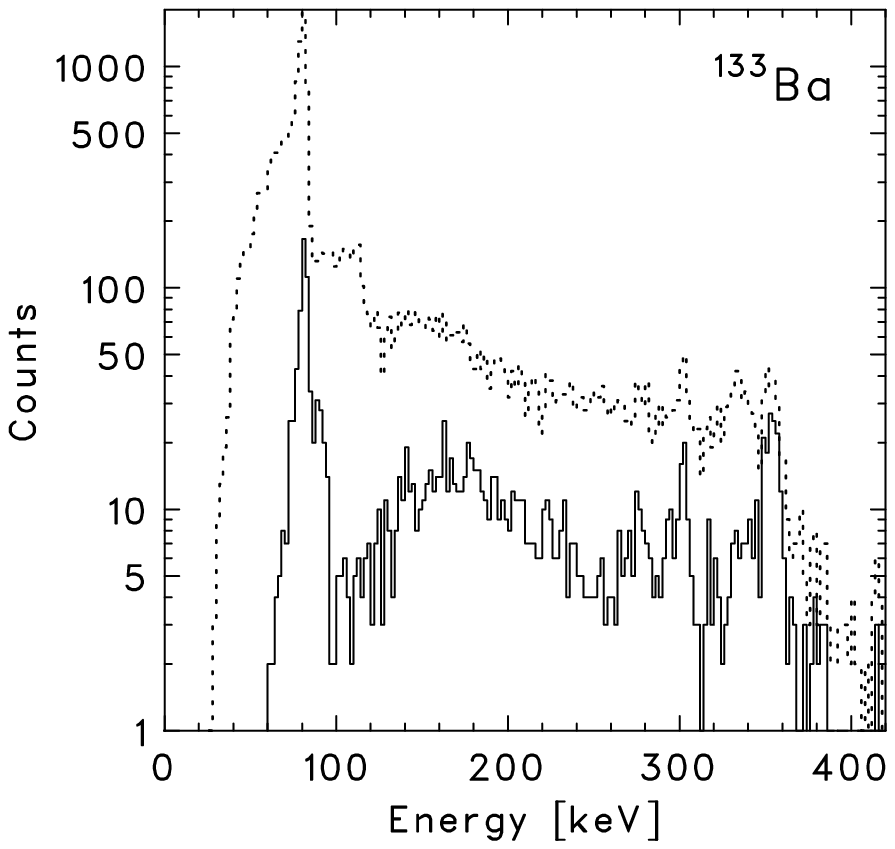}
\caption{Compton reconstructed spectra of ${}^{57}{\rm Co}$ (left) and 
${}^{133}{\rm Ba}$ (right). Dotted lines show the simple sum of two-hit events, 
while solid lines show the sum of events identified to be emitted from the source 
by Compton reconstruction. Note that there are many events which are photo-absorbed 
in CdTe followed by K-escape emission in the former, which is excluded in the latter, 
so that the peak counts also becomes smaller in the latter.}
\label{fig:comp_spec}
\end{center}
\end{figure}

Energy spectra of the incident gamma-rays are obtained by summing 
deposited energy at the DSSD and the CdTe detector 
as shown in equation (\ref{eq:comp_ene}). 
The dotted lines in Figure \ref{fig:comp_spec} show the spectra 
which we obtained by simply summing the deposited energy at each detector 
for all ``two-hit events''. In these spectra, the gamma-ray peaks from the 
sources can not be clearly identified since we can not distinguish ``one-Compton and 
absorption events'' from other ``two-hit events'', such events in which  
gamma-rays are scattered in the DSSD and scattered again in the CdTe and so on.   
Then we performed event selection by Compton kinematics. 
We selected the events satisfying 
$|\theta_{\rm Comp} - \theta_{\rm Geom}|< 16^{\circ}$. 
The solid lines in Figure \ref{fig:comp_spec} show the spectra of gamma-rays 
from ${}^{57}{\rm Co}$ and ${}^{133}{\rm Ba}$ after this selection. 
In these spectra, the gamma-ray peaks from the source are relatively enhanced. 

The energy resolution is 3.8~keV (FWHM) for 122~keV gamma-rays 
and 7.6~keV for 356~keV. These values are generally consistent with 
the energy resolution of the DSSD  (1.9~keV) and the CdTe pixel detectors 
(3.2~keV for 122~keV and 7.5~keV for 356~keV) under this setup, i.e. the 
energy resolution of this Compton telescope is expressed as 
\begin{eqnarray}
\Delta E = \sqrt{(\Delta E_{\rm Si})^2 + (\Delta E_{\rm CdTe})^2},
\end{eqnarray}
where $\Delta E_{\rm Si}$ and $\Delta E_{\rm CdTe}$ are the energy 
resolution of the DSSD and the CdTe detector, respectively. 

\subsection{Compton Reconstruction of K-escape Events}
As described above, significant number of 
fluorescence X-rays from the CdTe detectors are 
detected in the DSSD due to high fluorescence yield of 
Cd and Te ($\sim 80\%$ for the K shell). 
Depending on the detector geometry, K-escape events can share significant 
ratio in total events. Therefore, 
we attempted to reconstruct events in which photons are scattered at the 
DSSD, absorbed at the CdTe, K-escape takes place and the fluorescence 
line is detected again in the DSSD. 

\begin{figure}[htbp]
\begin{center}
\includegraphics[width=.4\linewidth,clip]{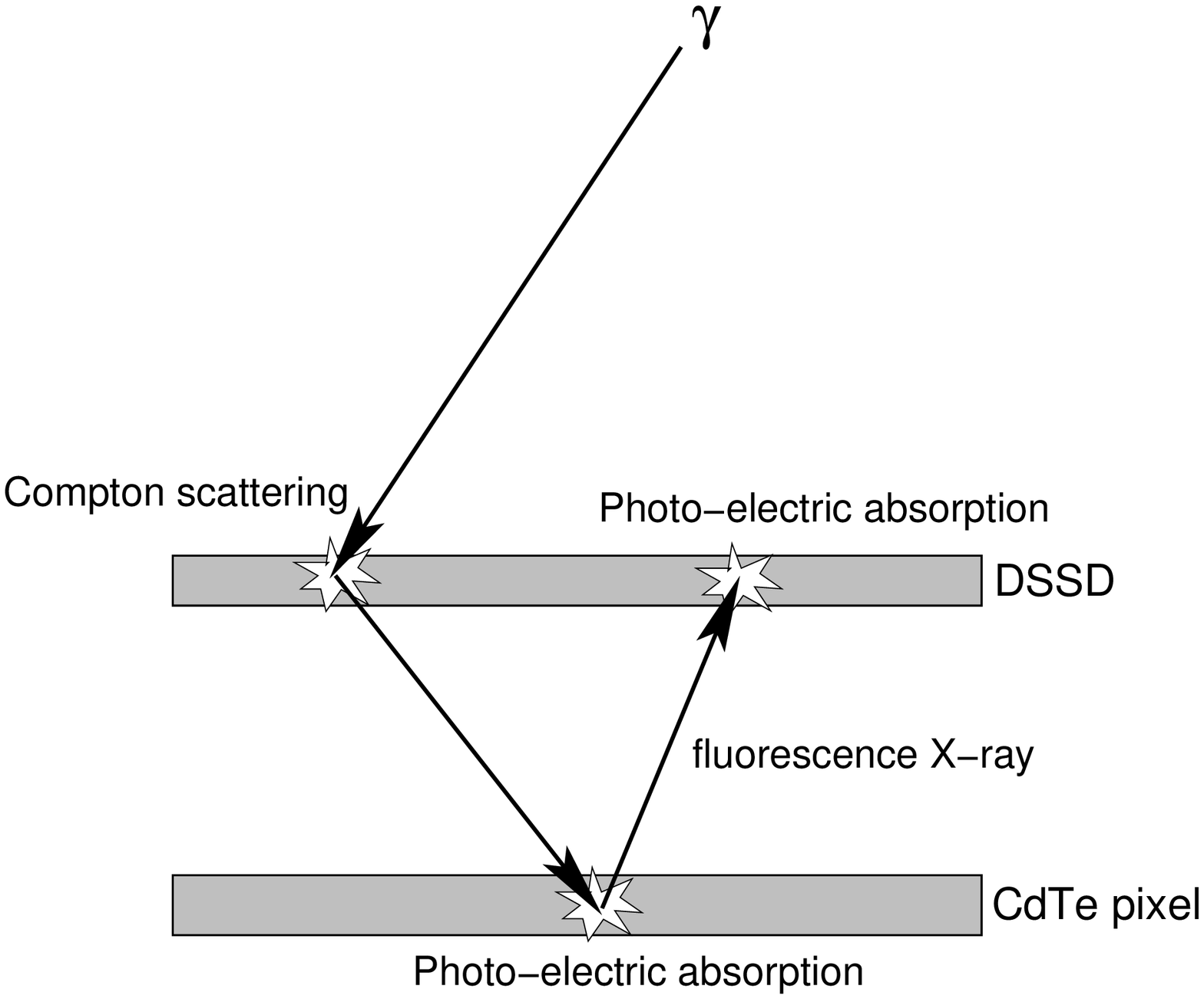}
\includegraphics[width=.45\linewidth,clip]{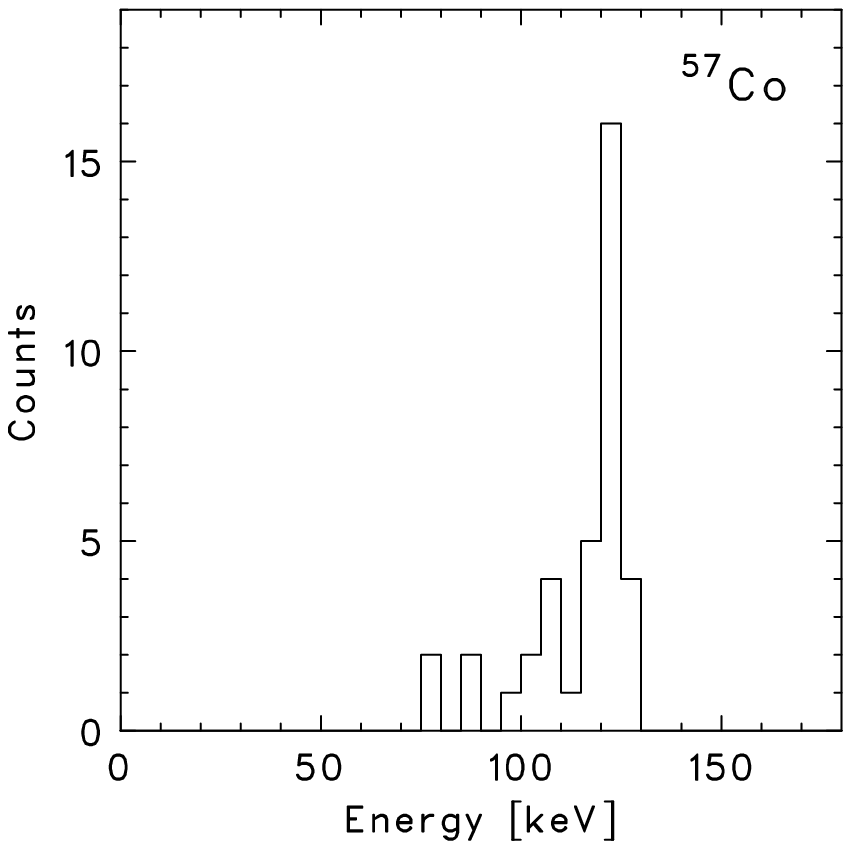}
\caption{Three-hit events by K-escape (left) and the reconstructed spectrum 
of ${}^{57}{\rm Co}$ obtained from such events (right).}
\label{fig:comp_spec_kx}
\end{center}
\end{figure}

First, we selected three-hit events, twice in the DSSD and once in the CdTe 
detector. Then we further select the events in which one of the deposited 
energy in the DSSD is between 20~keV and 34~keV. After adding these 
measured value to the energy observed in the CdTe detector, 
we calculated the Compton-scattering angle as 
the usual two-hit events. Figure \ref{fig:comp_spec_kx} (right) shows a 
spectrum of ${}^{57}{\rm Co}$ obtained from this analysis. Only the events 
which satisfy $|\theta_{\rm Comp} - \theta_{\rm Geom}|< 16^{\circ}$ 
are selected. The peak around 122~keV is clearly identified. 

\section{Implications from the prototype}
\begin{figure}[thbp]
\begin{center}
\includegraphics[width=.4\linewidth,clip]{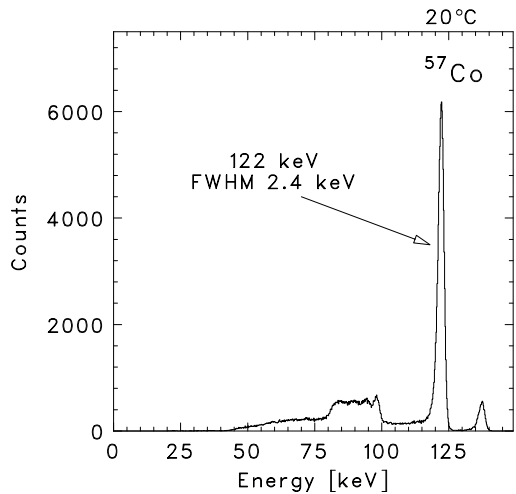}
\caption{${}^{57}{\rm Co}$ spectrum obtained with our latest version of the 
$8 \times 8$ CdTe pixel detector operated at room temperature. The applied bias 
voltage is 600~V.}
\label{fig:new_8x8_spec}
\end{center}
\end{figure}

From \S\ref{subsec:angle_res}, it is clear that 
the angular resolution of the prototype Compton telescope can be improved if 
the energy resolution becomes better especially for lower energy and if 
the position resolution becomes better for higher energy. 
In view of energy resolution, the CdTe detectors used in the prototype 
Compton telescope are not satisfactory. Actually, the best channel 
in other detector with the same design 
shows an energy resolution of 1.7~keV. 
We are currently trying to improve this properties 
by optimizing parameters for bump bonding. Figure \ref{fig:new_8x8_spec} shows 
the sum spectrum of all channels of the latest version of 
$8 \times 8$ CdTe pixel device, obtained at 
$20~{\rm {}^{\circ}C}$ with a bias voltage of 600~V. An 
energy resolution of 2.4~keV (FWHM) at 122~keV is obtained with the detector. 
Development of CdTe pixel detectors with smaller pixel size 
is also in underway\cite{Oonuki}. 
If we can make the pixel size smaller, 
we will improve not only the position resolution but also the energy resolution since 
the capacitance and the leakage current per pixel can be reduced. 
However, we need to be careful for charge sharing or K-escape to the other pixels 
in smaller pixel size detectors. 

Obtaining higher detection efficiency is another crucial issue for the next version 
of the Compton telescope. 
Currently, we are working on two directions. 
One is to make a multi-layer DSSD, up to 24 layers or more. 
Another way is to improve the thickness of the detector. 
We are now developing $500~{\rm \mu m}$ thick DSSD 
and thicker CdTe pixel detectors up to 5~mm. 
  
\section{Conclusion}
Si/CdTe semiconductor Compton telescope is attractive to explore the 
universe in sub-MeV gamma-ray region. We constructed a prototype Compton 
telescope which consists of one layer of DSSD and one layer of CdTe pixel detectors. 
With this prototype detector, we succeeded in Compton reconstruction of images 
and spectra in the energy band from 80~keV to 400~keV. The obtained angular 
resolution is $5.7^{\circ}$ and $9.9^{\circ}$ at 122~keV and 356~keV, respectively, 
and the energy resolution is 3.6~keV (FWHM) at 122~keV and 7.6~keV (FWHM) at 
356~keV. The lower energy coverage down to 80~keV owes to good energy 
resolution and lower energy threshold of our detectors. 
We also succeeded in reconstruction of three-hit events by Compton scattering 
and K-escape, which becomes significant under some configuration of Compton 
telescopes. 
With these results, we have experimentally demonstrated the high capabilities of 
a Si/CdTe Compton telescope.

\end{document}